\documentclass[%
 aip,
 amsmath,amssymb,
 reprint,%
]{revtex4-1}

\usepackage{graphicx}
\usepackage{dcolumn}
\usepackage{bm}

\usepackage[utf8]{inputenc}
\usepackage[T1]{fontenc}
\usepackage{mathptmx}
\usepackage{etoolbox}
\usepackage[usenames,dvipsnames]{color}

\makeatletter
\def\@email#1#2{%
 \endgroup
 \patchcmd{\titleblock@produce}
  {\frontmatter@RRAPformat}
  {\frontmatter@RRAPformat{\produce@RRAP{*#1\href{mailto:#2}{#2}}}\frontmatter@RRAPformat}
  {}{}
}%
\makeatother
\begin{document}

\preprint{AIP/123-QED}

\title[]{50 Ohm Transmission Lines with Extreme Wavelength Compression Based on Superconducting Nanowires on High-Permittivity Substrates}
\author{Daniel F. Santavicca}
 \email[]{daniel.santavicca@unf.edu}
 \affiliation{ 
Department of Physics, University of North Florida, Jacksonville, Florida 32224 USA}
\author{Marco Colangelo}
\affiliation{Department of Electrical Engineering and Computer Science, Massachusetts Institute of Technology,
Cambridge, Massachusetts 02139, USA}
\author{Carleigh R. Eagle}
\affiliation{ 
Department of Physics, University of North Florida, Jacksonville, Florida 32224 USA}
\author{Maitri P. Warusawithana}
\affiliation{ 
Department of Physics, University of North Florida, Jacksonville, Florida 32224 USA}
\author{Karl K. Berggren}
\affiliation{Department of Electrical Engineering and Computer Science, Massachusetts Institute of Technology,
Cambridge, Massachusetts 02139, USA}

\date{\today}

\begin{abstract}
We demonstrate impedance-matched low-loss transmission lines with a signal wavelength more than 150 times smaller than the free space wavelength using superconducting nanowires on high permittivity substrates. A niobium nitride thin film is patterned in a coplanar waveguide (CPW) transmission line geometry on a bilayer substrate consisting of 100 nm of epitaxial strontium titanate on high-resitivity silicon. The use of strontium titanate on silicon enables wafer-scale fabrication and maximizes process compatibility. It also makes it possible to realize a $50$ $\Omega$ characteristic impedance across a wide range of CPW widths, from the nanoscale to the macroscale. We fabricated and characterized an approximately $50$ $\Omega$ CPW device with two half-wave stub resonators. Comparing the measured transmission coefficient to numerical simulations, we determine that the strontium titanate film has a dielectric constant of $1.1 \times 10^3$ and a loss tangent of not more than 0.009. To facilitate the design of distributed microwave devices based on this type of material system, we describe an analytical model of the CPW properties that gives good agreement with both measurements and simulations. 
\end{abstract}

\maketitle

Cryogenic microwave circuitry has garnered increased interest in recent years, driven in part by the field of quantum computing based on superconducting circuits.\cite{Schoelkopf,Kjaergaard,Wendin} It is also important for the development of large-format arrays of cryogenic particle detectors\cite{Wollman} and high-speed classical computing using superconducting circuits.\cite{Tolpygo} The size of distributed microwave components -- such as filters, resonators, couplers, circulators, and travelling wave parametric amplifiers --  is limited by the signal wavelength, which is $\approx 4$ cm at 5 GHz in standard material systems. This large size makes on-chip integration difficult and is one of the major bottlenecks in scaling up cryogenic microwave systems.\cite{Gambetta,Hornibrook,Brecht} \par

For low-loss transmission lines, the characteristic impedance is $Z_0 = \sqrt{\mathcal{L}/\mathcal{C}}$ and the phase velocity is $v_p = 1/\sqrt{\mathcal{LC}}$, where $\mathcal{L}$ is the inductance per unit length and $\mathcal{C}$ is the capacitance per unit length. The ratio of the signal wavelength on the transmission line to the free space wavelength is the same as the ratio of the phase velocity on the transmission line to the speed of light in free space. Standard transmission lines, such as an RG-58 coaxial cable or a $50$ $\Omega$ microstrip on FR4 circuit board, have a signal wavelength that is approximately two thirds of the free space wavelength. The signal wavelength can be reduced while maintaining a matched impedance by proportionally increasing both $\mathcal{L}$ and $\mathcal{C}$.  Previous work has developed slow-wave transmission lines using conventional materials and increasing $\mathcal{L}$ and $\mathcal{C}$ by manipulating the transmission line geometry.\cite{Seki,Chang,Rosa} This approach has been successful in reducing the signal velocity and wavelength by up to an order of magnitude but not further.\par

Superconducting nanowires made from low carrier density materials can have a kinetic inductance that is two or more orders of magnitude larger than their magnetic inductance.\cite{Annunziata,Hazard,Pop} One such material is niobium nitride (NbN), which has demonstrated inductances per unit length $\sim 10^2 - 10^3$ pH/$\mu$m in a 100 nm wide, ultra-thin nanowire.\cite{Annunziata,Santavicca} This compares to a magnetic inductance of $\sim 1$ pH/$\mu$m. The large kinetic inductance naturally leads to a large characteristic impedance, $\sim k\Omega$, when patterned in a transmission line geometry.\cite{Santavicca,Colangelo} Such large impedances can be useful for decoupling from the electromagnetic environment but create challenges for coupling to standard microwave circuitry, which is commonly designed around a $50$ $\Omega$ impedance. Incorporating the nanowire in a high permittivity dielectric medium brings down the characteristic impedance and further shrinks the signal wavelength. High permittivity oxides such as hafnium dioxide (dielectric constant $\epsilon_r \approx 25$) and titantium dioxide ($\epsilon_r \approx 80$) offer significant enhancement over silicon dioxide ($\epsilon_r = 3.9$)\cite{Robertson} but do not have a large enough permittivity to offset the extremely large kinetic inductance of a NbN nanowire in order to achieve a $50$ $\Omega$ characteristic impedance.\par

Advances in materials science have allowed the synthesis of ceramic materials belonging to the class of perovskites with extremely high relative permittivities. One such material, strontium titanate (SrTiO\textsubscript{3}, abbreviated as STO), is a quantum paraelectric with an indirect bandgap of $3.25$ eV.\cite{Pai} It has been shown to have a relative permittivity exceeding $10^4$ in single crystal form at $4$ K, a value that decreases upon the application of an external electric field.\cite{Sakudo} This permittivity is sufficiently large to achieve a $50$ $\Omega$ characteristic impedance with a NbN nanowire. However, as the wire width increases, the kinetic inductance decreases, and the extreme permittivity of STO results in macroscale transmission lines with characteristic impedances well below $50$ $\Omega$.\cite{Davidovikj}\par 

In order to achieve a $50$ $\Omega$ characteristic impedance at both nanoscale and macroscale dimensions, we have utilized a thin layer of STO grown epitaxially on a bulk silicon (Si) wafer.\cite{Li,Warusawithana} At nanoscale dimensions, the STO dominates the effective permittivity, but as the transmission line width increases, the contribution of the STO to the effective permittivity decreases. The use of bulk Si also ensures the ability to fabricate on large-scale substrates, as single-crystal STO is generally not available in substrate sizes larger than two inches.\cite{Guguschev} Further, Si is a widely used substrate material, so its use increases process compatibility with other types of cryogenic and superconducting circuitry.\par   

Previous work described CPW resonators made from the high-temperature superconductor YBa\textsubscript{2}Cu\textsubscript{3}O\textsubscript{7-x} on substrates consisting of an STO film on bulk LaAlO\textsubscript{3}.\cite{Findikoglu,Adam} This work focused on utilizing the electric field-dependent polarizability of the STO to realize resonator tunability and the devices exhibited only modest compression of the signal wavelength by a factor of 2-3 compared to the free space value.\par 

\begin{figure}
\includegraphics[scale=0.31]{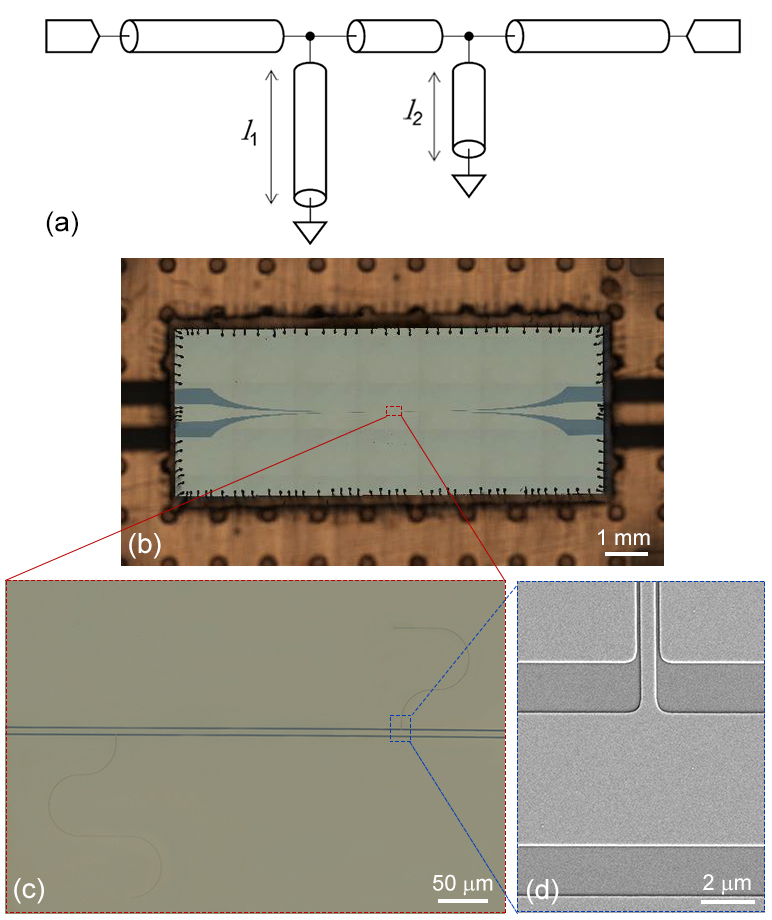}
\caption{\label{fig:Layout} (a) Equivalent circuit schematic of double-resonator device. The two half-wave stub resonators have lengths $l_1 = 300$ $\mu$m and $l_2 = 200$ $\mu$m. (b) Optical image of the full device chip mounted in a sample holder. The overall chip size is 10 mm $\times$ 4 mm. (c) Closeup of the center of the device chip showing the two resonators. (d) Scanning electron micrograph showing the connection of one resonator to the main CPW. The opposite end of the resonator connects to ground.}
\end{figure}

To demonstrate the potential for extreme wavelength compression using our material system, we fabricated the coplanar waveguide (CPW) double resonator device shown in figure \ref{fig:Layout}. The device maintains a characteristic impedance of approximately $50$ $\Omega$, with a CPW that transitions from a center conductor width of $400$ $\mu$m and a gap width between the center conductor and the coplanar grounds of $360$ $\mu$m at the edges of the chip -- large enough to facilitate wirebonding -- to a center conductor width of $5$ $\mu$m and a gap width of $1.75$ $\mu$m in the center of the chip. The transition maintains a constant fractional change in the center conductor width per unit length. In the center section, two half-wave stub resonators connect between the center conductor of the main CPW and ground. The resonators are in a CPW geometry with a center conductor width of $560$ nm and a gap width of $120$ nm. The length of the longer resonator is $300$ $\mu$m and the length of the shorter resonator is $200$ $\mu$m. An equivalent circuit model is shown in figure \ref{fig:Layout}a. \par 

The substrate consists of 100 nm of epitaxial (001) STO grown via molecular beam epitaxy (MBE) on a 370 $\mu$m thick high-resistivity (001) Si ($\rho > 1$ k$\Omega$cm). The initiation of the heteroepitaxial growth of STO on Si is key to obtaining a crystalline STO film\cite{Warusawithana} and involves a controlled sequence of steps that kinetically suppresses the formation of an amorphous oxide layer on Si\cite{Warusawithana,Li} and reduces the tendency for the STO film to form islands.\cite{Kourkoutis} Once a thin crystalline STO template was formed with in-plane epitaxial relationship STO [100] parallel to Si [110], the growth conditions were maintained at $\approx 700^{\circ}$C substrate temperature and $\approx 5 \times 10^{-7}$ Torr oxygen background pressure to obtain a 260 molecular-layer thick (100 nm) STO film. Throughout the growth, reflection high energy electron diffraction (RHEED) patterns showed good crystalline quality of the film.  See the Supplementary Material for additional growth details. Crystal quality was also evaluated using post-growth x-ray diffraction measurements: rocking curve measurements in $\omega$ of the out-of-plane STO 002 reflection showed a full-width at half-maximum of $\Delta \omega = 0.17^{\circ}$. To minimize oxygen vacancies which could lead to increased dielectric losses, the film was cooled down to below $150$ $^{\circ}$C under an oxygen background pressure of $\approx 5 \times 10^{-7}$ Torr before removing the sample from the MBE chamber.\par

A $\approx 15$ nm thick NbN film\cite{Medeiros} was deposited on the STO-Si substrate via reactive magnetron sputtering of niobium in the presence of nitrogen gas on the room temperature substrate. The device was patterned using 125 kV electron beam lithography with ZEP530A (Zeon corp.) positive tone resist. The patterns were developed in o-Xylene at $\approx 0$ $^{\circ}$C and were transferred into NbN with reactive ion etching in a CF\textsubscript{4} plasma. After etching, the resist residues were removed with N-Methyl-2-pyrrolidone. The measured critical temperature of the device was $12.0$ K and the switching current measured between one end of the CPW center conductor and ground -- i.e. measuring the two resonators in parallel -- was 813 $\mu$A at a temperature of 1.5 K.\par

The transmission coefficient $|S_{21}|^2 = 10\log_{10}\left(P_{out}/P_{in}\right)$ of this device measured at a bath temperature $T = 1.5$ K is shown in figure \ref{fig:Resonance}. A schematic of the experimental setup can be seen in the Supplementary Material. The signal power at the device is less than $-70$ dBm, ensuring that we do not excite non-linear effects. The data are normalized by measuring a sample with the same CPW geometry but no resonators. The fundamental half-wave resonances occur at 2.80 GHz and 4.20 GHz, corresponding to a signal wavelength on the resonators that is 178 times smaller than the free space wavelength.\par 

Also shown in figure \ref{fig:Resonance} is a simulation of the device using the AXIEM solver in AWR Microwave Office with an appropriately small mesh size. The dielectric anisotropy of STO is ignored in the simulation, and the superconducting NbN is modeled as a complex sheet impedance with a real part of zero and the imaginary part set to $2\pi f L_{K,sq}$ with $f$ the simulation frequency and $L_{K,sq}$ the sheet kinetic inductance.\cite{Santavicca} $L_{K,sq}$ can be determined from the normal-state sheet resistance $R_{sq}$ using
\begin{equation}
    L_{K,sq} = \frac{R_{sq}h}{2\pi^2\Delta}\frac{1}{\tanh{\left( \frac{\Delta}{2k_BT}\right)}}
\end{equation}
where $h$ is Planck's constant, $\Delta$ is the superconducting energy gap, and $k_B$ is Boltzmann's constant.\cite{Annunziata} In the limit that the temperature $T$ is much smaller than the superconducting critical temperature $T_C$, $\Delta \approx 1.76k_BT_C$ and the previous equation simplifies to $L_{K,sq} = R_{sq}h/\left(3.52\pi^2k_BT_C\right)$. In order to measure $R_{sq}$ without possible error due to substrate conductivity, a NbN film was deposited on a thermally-oxidized Si chip at the same time as the deposition on the STO-Si chip. This sample has $R_{sq} = 187$ $\Omega$ per square at $20$ K, corresponding to $L_{K,sq} = 21.5$ pH per square in the low temperature ($T \ll T_C$) limit, which is the value used in the simulation. 

The values of the STO dielectric constant and loss tangent are adjusted in the simulation to achieve the best agreement with the data. This leads to a dielectric constant of $1.1 \times 10^3$. The corresponding characteristic impedance of the resonator CPW is 66 $\Omega$, close to the target value of $50$ $\Omega$. The CPW gap size could be reduced to bring the impedance closer to the target value. While the STO permittivity is lower than values reported for bulk single crystals, it is comparable to previously reported values for thin-film STO,\cite{Li2,Xi} likely because additional disorder in the thin film material reduces the polarizability. The loss tangent determined from the simulation was $\tan \delta = 0.009$. We note that all other materials in the simulation were assumed to be completely lossless, and hence this value represents an upper bound on the actual loss tangent of the STO. Both resonances have a quality factor $Q = 160$ which is likely limited by material losses.\par 

\begin{figure}
\includegraphics[scale=0.6]{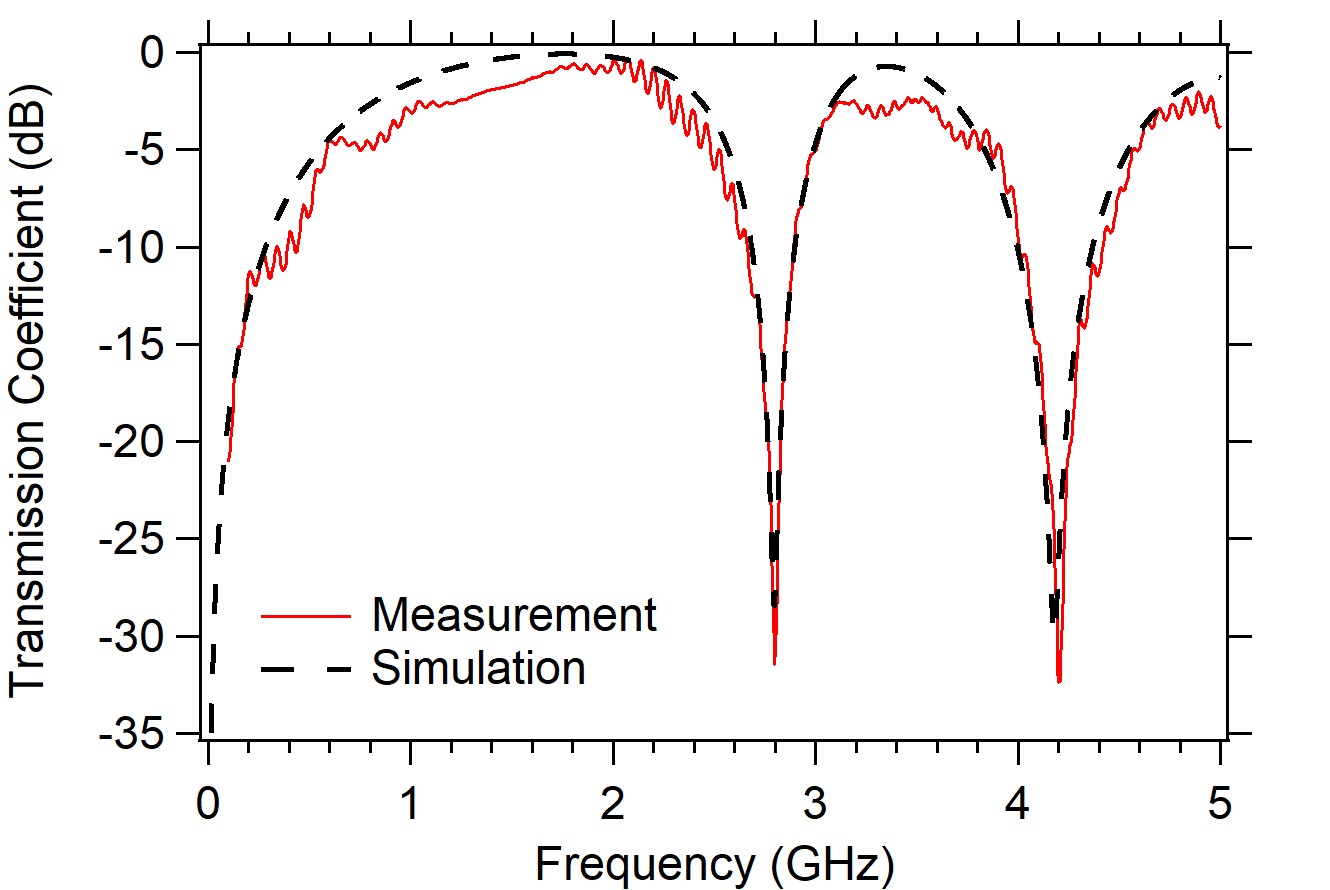}
\caption{\label{fig:Resonance} Measured and simulated transmission coefficient $|S_{21}|^2$ of the double-resonator device shown in figure 1. Data were taken at a temperature of $1.5$ K with zero bias current.}
\end{figure}

The demonstrated loss tangent is sufficiently low for many types of microwave devices but can likely be improved. We believe that the losses may be dominated by finite resistivity of the STO arising from oxygen vacancies. In future work, we plan to investigate further optimization of the STO growth to minimize its conductivity and maximize its permittivity. We also plan to investigate the dependence of the dielectric constant and loss tangent on temperature and electric field.\par 

When performing electromagnetic device simulations, generally one wants to ensure that the simulation mesh is sufficiently small such that further reduction in the mesh size does not change the simulation results. For this material system, we found that satisfying this condition required a considerably smaller mesh size in the vicinity of the CPW gap than is needed for conventional materials. (Further details are provided in the Supplementary Material.) As a result of the fine mesh size, the simulation time and memory requirement become considerable; the simulation results shown in figure \ref{fig:Resonance} took more than 24 hours.\par 

Such a long simulation time poses challenges for iterative device design and optimization. In order to facilitate more rapid device design, we developed an analytical model for CPW transmission lines using this material system. This analytical model uses the established conformal mapping approach for CPW transmission lines made from conventional materials\cite{Gevorgian,Garg} combined with a model based on BCS theory developed by Clem for the kinetic inductance per unit length of a superconducting CPW.\cite{Clem}\par 

First the capacitance per unit length $\mathcal{C}$ of the CPW geometry is calculated using a Schwarz-Christoffel conformal mapping. This process has been described previously\cite{Gevorgian,Garg} and the details are in the Supplementary Material. While single-crystal STO is an anisotropic dielectric, with the largest low-temperature polarizability in the direction of the (110) crystal axis,\cite{Sakudo} this calculation ignores the anisotropy. 

The total inductance per unit length $\mathcal{L}$ is the sum of the magnetic inductance per unit length $\mathcal{L}_{M}$ and the kinetic inductance per unit length $\mathcal{L}_{K}$. Considering only magnetic inductance, $v_P = c/\sqrt{\epsilon_{eff}}$ where $c$ is the speed of light in free space and $\epsilon_{eff}$ is the effective dielectric constant (calculated as described in the Supplementary Material), and hence $\mathcal{L}_{M}$ can be found from: 
\begin{equation}
    \mathcal{L}_M = \frac{\epsilon_{eff}}{\mathcal{C}c^2}
\end{equation} \par
For a superconductor with thickness $d < 2\lambda_L$, where $\lambda_L$ is the London penetration depth, the distribution of the current across the width of the wire is no longer governed by $\lambda_L$ but rather by the Pearl length $\Lambda_P = 2\lambda_L^2/d$. This is the case for our samples. When the center conductor width $2s \ll \Lambda_P$, the current distribution in the center conductor is approximately uniform and the kinetic inductance per unit length $\mathcal{L}_K$ is, to a good approximation, equal to the sheet kinetic inductance divided by $2s$. The sheet kinetic inductance $L_{K,sq}$ is found as described previously, with a value of $21.5$ pH per square in the low temperature ($T \ll T_C$) limit. $\lambda_L$ can then be calculated from 
\begin{equation}
    L_{K,sq} = \frac{\mu_0\lambda_L^2}{d}
\end{equation}
where $\mu_0$ is the permeability of free space. This yields $\lambda_L = 507$ nm and $\Lambda_P = 34.2$ $\mu$m.
\par
A procedure for calculating $\mathcal{L}_K$ for all center conductor widths was developed by Clem.\cite{Clem} In this approach, $\mathcal{L}_K$ is found from
\begin{equation}
    \mathcal{L}_K = \frac{\mu_0\Lambda_P}{4s}f(k,p)
\end{equation}
where 
\begin{equation}
    f(k,p) = \frac{(k+p^2)\tanh^{-1}(p)-(1+kp^2)\tanh^{-1}(kp)}{p(1-k^2)\left( \tanh^{-1}(p)\right)^2}
\end{equation}
with $k = s/(s+g)$ where $2s$ is the CPW center conductor width and $g$ is the gap width between the center conductor and ground. The parameter $p$ is calculated as described in Clem\cite{Clem} but has the following useful approximations:
\begin{equation}
    p \approx 0.63/\sqrt{\Lambda_P/s} \quad \textrm{for} \quad \Lambda_P \gg s
\end{equation}
\begin{equation}
    p \approx 1 - 0.67\Lambda_P/s \quad \textrm{for} \quad \Lambda_P \ll s
\end{equation}
This approach to calculating $\mathcal{L}_K$ gives results that are very close to $\mathcal{L}_K = L_{K,sq}/(2s)$ when $\Lambda_P \gg s$. \par

\begin{figure}
\includegraphics[scale=0.6]{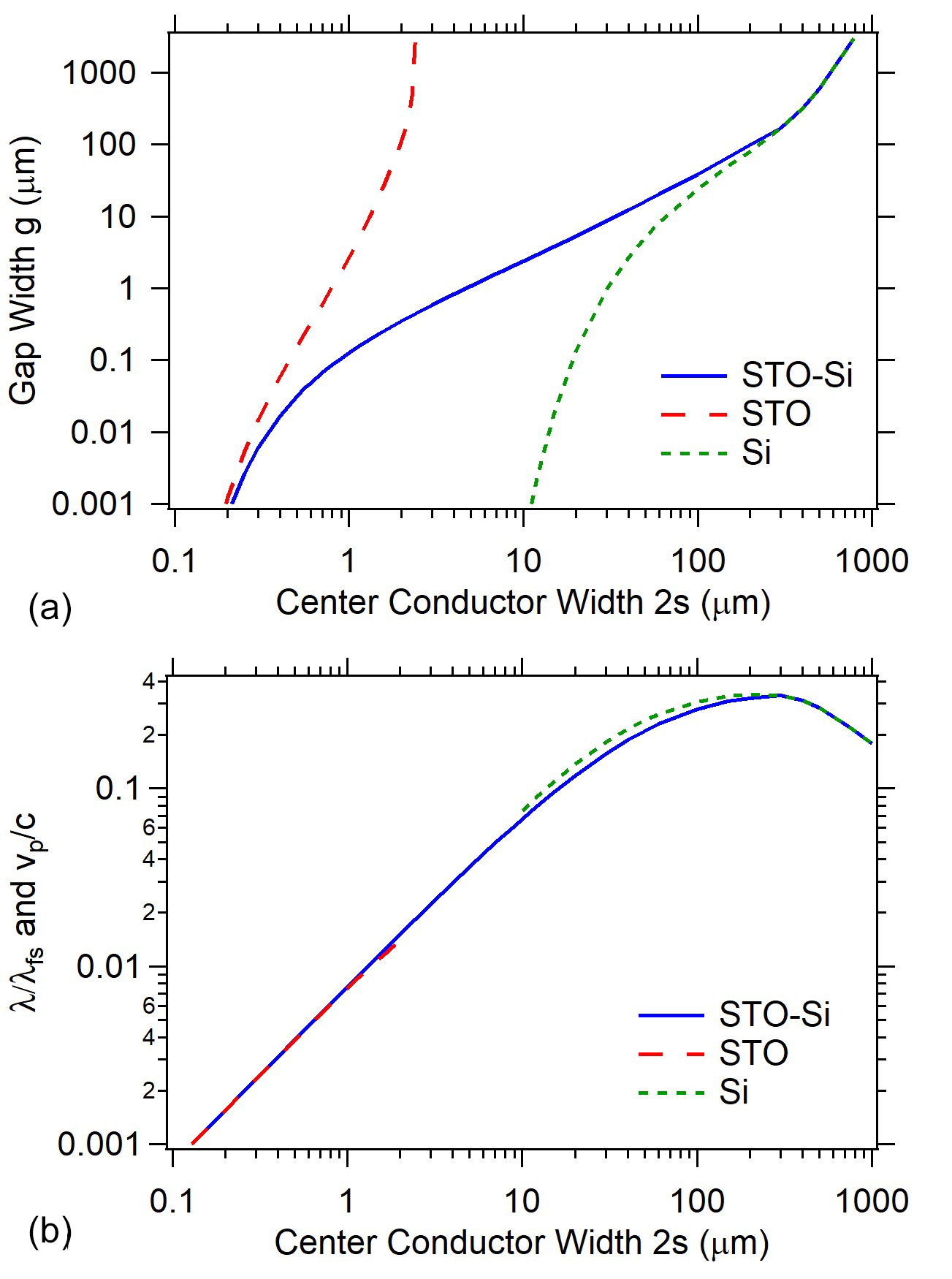}
\caption{\label{fig:Analytical} (a) Analytical calculation of the CPW gap width $g$ required to achieve a characteristic impedance of 50 $\Omega$ as a function of the center conductor width $2s$. Results are shown for the bilayer substrate consisting of $100$ nm STO (dielectric constant $= 1.1 \times 10^3$) on top of $370$ $\mu$m Si (dielectric constant $= 11.7$). Also shown for comparison are the results for a substrate of the same total thickness consisting entirely of STO or Si. (b) Corresponding ratio of the signal wavelength $\lambda$ to the free space wavelength $\lambda_{fs}$, which is also the ratio of the phase velocity $v_p$ to the speed of light in free space $c$.}
\end{figure}

Using this analytical approach with the sample parameters summarized in Table 1, we calculate the value of the gap width $g$ that gives a characteristic impedance $Z_0 = 50$ $\Omega$ as a function of the center conductor width $2s$. The results of this calculation are shown in figure \ref{fig:Analytical}. For comparison, we also show the results for the case with an STO substrate of the same total thickness as well as the case with an Si substrate of the same total thickness. We see that the STO-Si bilayer substrate allows us to achieve a $50$ $\Omega$ impedance over a much wider range of center conductor widths than either a STO or Si substrate. We found agreement to better than $5$\% between the numerical simulations and the analytical calculations provided that a sufficiently fine mesh size was used in the vicinity of the CPW gap. Because of the long times required for accurate simulations, this analytical model can be very useful for initial device design. \par 

\begin{table}
\caption{\label{tab:table}Summary of sample parameters used in the analytical CPW model}
\begin{ruledtabular}
\begin{tabular}{lcr}
Parameter & Value\\
\hline
NbN thickness & $15$ nm\\
NbN normal state sheet resistance & $187$ $\Omega$/sq \\
NbN critical temperature & 12.0 K \\
NbN sheet kinetic inductance ($T \ll T_C$) & $21.5$ pH/sq\\
NbN London penetration depth ($T \ll T_C$) & $507$ nm \\
NbN Pearl length ($T \ll T_C$) & $34.2$ $\mu$m\\
STO relative permittivity & $1.1 \times 10^3$\\
STO thickness & $100$ nm \\
Si relative permittivity & $11.7$ \\
Si thickness & $370$ $\mu$m \\
\end{tabular}
\end{ruledtabular}
\end{table}

Devices that combine high kinetic inductance superconducting thin films and high permittivity substrates enable the creation of impedance-matched microwave devices in which the signal wavelength and velocity can be reduced by more than two orders of magnitude, as demonstrated by the CPW double-resonator device described in this paper. This extreme wavelength compression facilitates the on-chip integration of many types of distributed microwave devices that would be prohibitively large using conventional material systems, expanding the possibilities for scaling up cryogenic microwave systems.\\ \par

See the supplementary material for additional details on the STO growth, the experimental setup for device characterization, numerical simulations, and the analytical model.

\begin{acknowledgments}
This work was supported by National Science Foundation grants ECCS-2000778 (UNF) and ECCS-2000743 (MIT). M.C. acknowledges support from the Claude E. Shannon Award.
\end{acknowledgments}

\bibliography{aipsamp}

\end{document}